\documentclass{article}

\def\starcom {\stackrel{\star}{,}}
\def\lam {\Lambda}

\def\t{\theta}
\def\D {\Delta}
\def\l{\lambda}
\def\L {\Lambda}

\def\del{\partial}
\def\ha{\frac{1}{2}}
\def\be{\begin{equation}}
\def\ee{\end{equation}}
\def\ba{\begin{eqnarray}}
\def\ea{\end{eqnarray}}
\begin{document}
\begin{titlepage}

\vskip -0.55cm 
\begin{center}

\vskip .15in

\renewcommand{\thefootnote}{\fnsymbol{footnote}}
{\large \bf The Seiberg-Witten Map for Noncommutative Gauge Theories}        
\vskip .25in
B. L. Cerchiai\footnote{email address: BLCerchiai@lbl.gov}, 
A. F. Pasqua\footnote{email address: pasqua@socrates.berkeley.edu},
B. Zumino\footnote{email address: zumino@thsrv.lbl.gov}
\vskip .25in

{\em    Department of Physics  \\
        University of California   \\
                                and     \\
        Theoretical Physics Group   \\
        Lawrence Berkeley National Laboratory  \\
        University of California   \\
        Berkeley, California 94720}
\end{center}
\vskip .25in

\begin{abstract}
The Seiberg-Witten map for noncommutative Yang-Mills theories is studied and
methods for its explicit construction are discussed which are
valid for any gauge group. In particular the use of the evolution equation
is described in some detail and its relation to the cohomological approach is
elucidated. Cohomological methods which are applicable to gauge theories
requiring the Batalin-Vilkoviskii antifield formalism are briefly mentioned.
Also, the analogy of the Weyl-Moyal star product with the star product of
open bosonic string field theory and possible ramifications of this analogy
are briefly mentioned.
\end{abstract}
\end{titlepage}
\setcounter{footnote}{0}

\section{Introduction}

Noncommutative field theories have recently received much attention.
Seiberg and Witten \cite{SW} have argued that certain noncommutative 
gauge theories are equivalent to commutative ones and in particular that 
there exists a map from a commutative gauge field to a noncommutative one,
which is compatible with the gauge structure of each. 
This map has become known as the Seiberg-Witten (SW) map.

In two recent papers \cite{BCPVZ,BCZ} we have discussed a cohomological
method for constructing explicitly this map. Here we describe a slightly
modified procedure based on the idea
that the structure equations of the gauge group of the noncommutative
theory are a deformation of those of the gauge group of the commutative theory.
We will consider gauge theories on the noncommutative space defined by
\be
\left [x^i \starcom x^j \right ] =i\theta ^{ij}~,
\ee 
where $\theta$ is a constant Poisson tensor. The  ``$\star$'' 
operation is the associative Weyl-Moyal product
\be
\label{WMproduct}
f \star g= f\, e^{\frac{i}{2}\theta ^{ij}\stackrel{\leftarrow}{\partial _i}
\stackrel{\rightarrow}{\partial_j}}g~.
\ee
We believe that our methods are much more general, and can in fact be
used even when $\theta$ is not constant, but in this paper we shall make use 
of the fact that the $x^i$ derivative $\del_i$ of functions satisfies the
Leibniz rule with respect to the star product
\be
\del_i(f \star g)=(\del_i f) \star g+f \star (\del_i g),
\ee
just as it does with respect to the ordinary product. This simple relation
requires $\t$ to be constant.

\section{Structure Equations}

The structure equations of a gauge group can be expressed in terms of a
ghost field $\lambda(x)$ and the gauge potential $a_i(x)$ by giving the
action of the BRST operator $s$
\ba
s \l&=&i \l \cdot \l, \label{unstrl} \\
s a_i&=&\del_i \l-i a_i \cdot \l+i \l \cdot a_i~.\label{unstra}
\ea
Here $\l$ and $a_i$ are valued in a Lie algebra and can be represented by
matrices, the matrix elements of the ghost field being anticommuting functions
of $x$. In a representation the product would imply matrix multiplication.
The operator $s$ is an odd superderivation of ghost number one
\ba
s (f \cdot g)&=& (sf) \cdot g \pm f \cdot sg, \label{leibniz} \\
s^2&=&0, \label{nil}
\ea
which commutes with the derivatives
\be
s \del_i=\del_i s.
\label{sder}
\ee
As usual, the signs in (\ref{leibniz}) depend on the parity of $f$.
Our task is to deform the above structure equations into
\ba
s \L &=& i \L \star \L, \label{strl}  \\
s A_i&=&\del_i \L -i [ A_i \starcom \L ], \label{stra}
\ea
where $A_i=A_i(a,\del a,\del^2 a, \ldots)$ is an even local functional of 
$a_i$, of ghost number zero, and $\L=\L(\l,\del\l,\ldots,a,\del a,\ldots)$
is an odd local functional of $a_i$ and $\l$, of ghost number one 
(like $\l$). We take $s$ to be undeformed and to satisfy (\ref{nil}),
(\ref{sder}) and
\be
s (f \star g)=sf \star g\pm f \star sg.
\ee
The solution consists in finding explicit expressions for the functionals
$A_i$ and $\L$. This can be done as expansions in $\t$
\ba
\L&=&\L^{(0)}+\L^{(1)}+\ldots, \quad \Lambda^{(0)}=\l~, \\
A_i&=&A_i^{(0)}+A_i^{(1)}+\ldots, \quad A_i^{(0)}=a_i~.
\ea
The first order terms were given already in \cite{SW}
\ba
\L^{(1)}&=&\frac{1}{4}\t^{kl} \left\{\del_k \l, a_l \right\} , 
\label{soll} \\
A_i^{(1)}&=&-\frac{1}{4}\t^{kl} \left\{a_k, \del_l a_i+ f_{li} \right\},
\label{sola}
\ea
where
\be
f_{li}=\del_l a_i-\del_i a_l-i[a_l,a_i]
\label{f}
\ee
is the commutative field strength, and expressions for $\L^{(2)}$ and 
$A_i^{(2)}$ are known~\cite{BCPVZ,GoHa,JMSSW}, see also below.

A systematic way to obtain the expansion in $\t$ was described
in \cite{BCPVZ,BCZ} and the consistency of the procedure was demonstrated
in \cite{PQS}. Each order in $\theta$ is manifestly local.

One sees already from (\ref{soll}) and (\ref{sola}) that $\L$ and $A_i$ 
cannot be Lie algebra valued in general, and we follow \cite{JSSW,JMSSW}
by allowing them to be in the enveloping algebra of the Lie algebra of $\l$
and $a_i$.
A representation of this Lie algebra lifts naturally to a representation of
its enveloping algebra.

\section{Evolution Equations}

There is an alternative approach for the study of the SW map, which is based on
a differential equation \cite{SW}. 
Let us introduce a ``time'' parameter $t$ in front of $\t$, in such a way that 
$\t^{ij} \rightarrow t \, \t^{ij}$, $\L \rightarrow \L(t)$ 
and $A_i \rightarrow A_i(t)$, while keeping $s$ independent of $t$.
Notice that $\L$ and $A_i$ acquire a $t$-dependence through $\t$.
Differentiating the structure equations (\ref{strl}) and (\ref{stra})
with respect to $t$, we obtain\footnote{As customary the dot denotes
differentiation with respect to $t$.}
\ba
s \stackrel{\scriptscriptstyle \bullet}{\Lambda}&=&i 
\stackrel{\scriptscriptstyle \bullet}{\L} \star \; \L +i \L \; \star 
\stackrel{\scriptscriptstyle \bullet}{\L} +i \L 
\stackrel{\scriptscriptstyle \bullet}{\star} \L, \label{timetl} \\
s \stackrel{\scriptscriptstyle \bullet}{A_i}&=&
\stackrel{\scriptscriptstyle \bullet}{A_i} \star \; \L + \L \; \star 
\stackrel{\scriptscriptstyle \bullet}{A_i} + D_i 
\stackrel{\scriptscriptstyle \bullet}{\Lambda}-i 
A_i \stackrel{\scriptscriptstyle \bullet}{\star} \L+i \L 
\stackrel{\scriptscriptstyle \bullet}{\star} A_i,
\label{timeta}
\ea
where
\be
D_i=\del_i -i[A_i \starcom \cdot ~]~
\ee
is the covariant derivative at time $t$.
The star product itself depends on the evolution parameter $t$, and therefore
it has also to be differentiated
\be
\star=e^{\ha i t \stackrel{\leftarrow}{\del_i} \t^{ij} 
\stackrel{\rightarrow}{\del_j}},
\qquad \stackrel{\scriptscriptstyle \bullet}{\star}=e^{\ha i t
\stackrel{\leftarrow}{\del_i} 
\t^{ij} \stackrel{\rightarrow}{\del_j}}\frac{i}{2} \stackrel{\leftarrow}{\del_k} 
\t^{kl} \stackrel{\rightarrow}{\del_l}.
\ee
Explicitly this yields
\be
f\stackrel{\scriptscriptstyle \bullet}{\star} g=i \frac{\t^{kl}}{2} \; 
\del_k f \star \del_l g.
\ee
Notice that for simplicity we have restricted ourselves to a linear path
in $\t$-space, i.e. we are considering a linear one-parameter family of 
deformations of $\t$. In principle it would be possible to consider an 
arbitrary variation with respect to $\t$ corresponding to
an arbitrary path in $\t$-space, like e.g. in \cite{GoHa}.

The structure of the right hand side of (\ref{timetl}) and (\ref{timeta}) 
leads in a natural way to the definition of a new
operator\footnote{$\Delta_t$ is a simple generalization of the operator
$\Delta$ introduced in \cite{BCPVZ}, which now should be called $\Delta_0$.
Also, what was called $\hat \Delta$ in \cite{BCZ} should now be called
$\Delta_1$.}
at time $t$:
\be
\Delta_t=\left \{\begin{array}{ll}
s -i \{\Lambda \starcom \cdot \} & 
\textrm{on odd quantities}, \\ \\
s -i [\Lambda \starcom \cdot ] &  
\textrm{on even quantities}.\\
\end{array}
\right.
\label{deltat}
\ee
It has the following properties
\ba
\Delta_t A_i=\partial_i \Lambda, ~~~ \Delta_t^2=0, ~~~
{[\Delta_t,D_i ]} = 0,\\
\Delta_t (f_1 \star f_2)=(\Delta_t f_1) \star f_2 \pm f_1 \star (\Delta_t f_2),
\ea
i.e. $\Delta_t$ is nilpotent, it commutes with the covariant derivative
at time $t$ and it satisfies a super-Leibniz rule. This is a consequence of the
fact that
\be
s^2=0, \qquad s \del_i=\del_i s,
\ee
and of the associativity of the star product.
Therefore, $\Delta_t$ can be interpreted as a coboundary operator in a suitably
defined cohomology.

Using the operators $\Delta_t$ and $D_i$ the equations (\ref{timetl}) and 
(\ref{timeta}) can be rewritten as
\ba
\Delta_t \stackrel{\scriptscriptstyle \bullet}{\Lambda}&=&-\ha \theta^{kl} 
\del_k \Lambda \star 
\del_l \Lambda = -\ha \theta^{kl} B_k \star B_l,  \label{swtl} \\
\Delta_t \stackrel{\scriptscriptstyle \bullet}{A_i}&=&D_i 
\stackrel{\scriptscriptstyle \bullet}{\Lambda}+\ha 
\theta^{kl} \{\del_k A_i \starcom \del_l \Lambda\}=
D_i \stackrel{\scriptscriptstyle \bullet}{\Lambda}+\ha 
\theta^{kl} \{\del_k A_i \starcom B_l \}. \label{swta}
\ea
Here we have introduced the notation 
\be
B_i=\del_i \L,
\label{defb}
\ee
which is useful because only derivatives of $\L$ enter in the right 
hand side of (\ref{swtl}) and (\ref{swta}), but never $\L$ itself.
The action of $\Delta_t$ in terms of these new variables $A_k$ and $B_k$ 
takes a particularly simple form 
\be
\Delta_t A_k=B_k, \qquad \Delta_t B_k=0.
\label{defdab}
\ee
With this action the consistency condition that $\Delta_t$ applied to
the right hand side of equation (\ref{swtl}) vanishes is verified. 
For (\ref{swta}) we find that $\Delta_t$ on the right hand side
gives $\ha \t^{kl} [ \Delta_t F_{ki} \starcom B_l]$. We will comment
on this later in section 4.

The differential evolution equations which provide a solution to the 
equations (\ref{swtl}), (\ref{swta}) are given by \cite{SW}
\ba
\stackrel{\scriptscriptstyle \bullet}{\Lambda}&=&\frac{1}{4}\t^{ij} 
\left\{\del_i \Lambda \starcom A_j \right\},
\label{l1} \\
 \stackrel{\scriptscriptstyle \bullet}{A_i}&=&-\frac{1}{4}\t^{kl}
\left\{A_k \starcom \del_l A_i+F_{li} \right\},
\label{a1}
\ea
where
\be
F_{li}=\del_l A_i - \del_i A_l - i [A_l \starcom A_i ]
\label{F}
\ee
is the noncommutative field strength.
This can be easily checked by substituting these expressions in (\ref{swtl})
and (\ref{swta}).

\section{The Homotopy Operator}

There is a way of computing the expressions (\ref{l1}) for 
$\stackrel{\scriptscriptstyle \bullet}{\Lambda}$ and (\ref{a1}) for 
$\stackrel{\scriptscriptstyle \bullet}{A_i}$ 
through a suitably defined homotopy operator $K_t$. 
Clearly, it is not possible to invert $\Delta_t$,
because it is nilpotent, but if we construct an operator such that
\be
K_t \Delta_t + \Delta_t K_t=1,
\label{kd}
\ee
then an equation of the form
\be
\Delta_t f=m,
\ee
with
\be
\Delta_t m=0,
\ee
has a solution of the type
\be
\label{homsol}
f=K_t m,
\ee
because
\be
m=\Delta_t K_t m +K_t \Delta_t m =\Delta_t K_t m.
\ee
The solution (\ref{homsol}) is not unique: $K_t m +\Delta_t h$, with some
appropriate $h$, is also a solution, since $\Delta_t^2=0$.
This is the same method we applied for $t=0$ in \cite{BCPVZ,BCZ}, which
closely follows the ideas developed in \cite{Zu} to study anomalies in
chiral gauge theories.

Let us construct such a homotopy operator $K_t$ explicitly. We start by
defining a linear operator $\tilde K_t$ such that
\be
\tilde K_t B_k=A_k, \qquad \tilde K_t A_k=0.
\ee
On both $A_k$ and $B_k$ it satisfies
\be
\tilde K_t \D_t+\D_t \tilde K_t=1.
\label{tkd}
\ee
Further, we require that it is a super-derivation
\be
\tilde K_t (f_1 \star f_2)=(\tilde K_t f_1) \star f_2 \pm f_1 \star 
(\tilde K_t f_2)
\ee
and that it commutes with $D_i$ and anticommutes with $s$
\be
[\tilde K_t, D_i]=0, \qquad \{\tilde K_t, s\}=0.
\ee

Notice that due to (\ref{tkd}) $\tilde K_t$ has to be odd and it decreases
the ghost number by one. Moreover, it is nilpotent on $A_i$, $B_i$
\be
\tilde K_t^2=0.
\label{ntk}
\ee

On monomials of higher order in $A_k$ and $B_k$, the homotopy operator $K_t$ 
cannot satisfy the Leibniz rule. If $d$ is the total order of such a monomial
$m$, then the action of $K_t$ on it has to be defined as
\be
K_t m=d^{-1} \tilde K_t m.
\ee
It is extended to general polynomials by linearity.  
Then $K_t$ satisfies (\ref{kd}) and from (\ref{ntk}) it follows that
\be
K_t^2=0.
\ee
Now, we can use $K_t$ to recover the solutions (\ref{l1}), (\ref{a1}) of the 
equations (\ref{swtl}), (\ref{swta}).
For $\L$ this is straightforward. We apply $K_t$ to the right hand side of 
(\ref{swtl}) and we get
\ba
\stackrel{\scriptscriptstyle \bullet}{\Lambda} &=&K_t \left(-\ha \theta^{kl} 
B_k \star B_l\right)=-\frac{1}{4} \theta^{kl}
\left(\tilde K_t B_k \star B_l-B_k \star \tilde K_t B_l\right)\\
&=&-\frac{1}{4} \theta^{kl}\left( A_k \star B_l -B_k \star A_l \right)=
\frac{1}{4} \t^{kl} \left\{B_k \star A_l \right\}, \nonumber
\ea
which coincides with (\ref{l1}).

For the gauge potential, however, there is a complication.
If we apply $\Delta_t$ to the right hand side of (\ref{swta}) we obtain
\be
\Delta_t \left( D_i \stackrel{\scriptscriptstyle \bullet}{\Lambda}+\ha 
\theta^{kl} \{\del_k A_i \starcom B_l \} \right)=\ha \t^{kl} 
[ \Delta_t F_{ki} \starcom B_l],
\ee
where
$\Delta_t F_{ki}=D_k B_i-D_i B_k+i[B_k,A_i]+i[A_k,B_i]$.
This expression vanishes only if we impose the condition that
\be
\Delta_t F_{ki}=0.
\label{constraint}
\ee
This property is true if we explicitly use the definition (\ref{defb}) of 
$B_i=\del_i \L$, but it has to be set as an additional constraint in the 
algebra generated by the $A_i$, the $B_i$
and their derivatives. In other words, such an algebra is not free. 
The homotopy operator $K_t$ can be defined only on $B$.
In order to solve this problem, we can add to the right hand side of 
(\ref{swta}) a term which is zero by the constraint (\ref{constraint}), but 
which makes the $\Delta_t$ of it vanish algebraically.
For this purpose we can choose e.g. $\ha \t^{kl} \{ \Delta_t F_{ki} 
\starcom A_l\}$ and consider the expression 
\be
U_i \equiv D_i \stackrel{\scriptscriptstyle \bullet}{\Lambda}+\ha \theta^{kl} 
(\{\del_k A_i \starcom B_l \} + \{ \Delta_t F_{ki} \starcom A_l\} ).
\ee
Then
\be
\Delta_t U_i=0
\ee
algebraically and we can apply the homotopy operator to $U_i$ and obtain 
(\ref{a1}).
This is the same procedure we have proposed in \cite{BCPVZ} and \cite{BCZ} 
to treat the analogous difficulty.

\section{Solutions to higher order in $\t$}

Observe that we can recover the first order in the $\t$ expansion as
\ba
\L^{(1)}&=&\stackrel{\scriptscriptstyle \bullet}{\Lambda}(t) \mid_{t=0}
=\frac{1}{4}\t^{ij} \left\{\del_i \lambda, a_j \right\} , \\
A_i^{(1)}&=&\stackrel{\scriptscriptstyle \bullet}{A_i}(t) \mid_{t=0}
=-\frac{1}{4}\t^{kl} \left\{a_k, \del_l a_i+ f_{li} \right\},
\ea
which yields the well-known solution found by Seiberg and Witten \cite{SW}.
More in general, once we have the expressions (\ref{l1}) and (\ref{a1}) 
to first order, the evolution equations provide a useful 
method for computing the terms of higher order in $\theta$ by just 
noticing that
\be
\L^{(n)}=\frac{1}{n!} {\del^n \L(t) \over \del t^n} \mid_{t=0}, \qquad
A_i^{(n)}=\frac{1}{n!} {\del^n A_i(t) \over \del t^n} \mid_{t=0}.
\ee
Therefore, by simply differentiating with respect to $t$, we can compute 
$\L^{(n)}$ and $A_i^{(n)}$. This is an alternative and easier technique
than applying the homotopy operator order by order as suggested 
in \cite{BCPVZ,BCZ}.

In particular to second order we get
\ba
\stackrel{\scriptscriptstyle \bullet \scriptscriptstyle \bullet}{\Lambda} 
&=& \frac{1}{4}\t^{kl}\Big( \left\{\del_k
\stackrel{\scriptscriptstyle \bullet}{\Lambda} \starcom A_l \right\}
+\left\{\del_k \L \starcom \stackrel{\scriptscriptstyle \bullet}{A_l} \right\}
+\del_k \L \stackrel{\scriptscriptstyle \bullet}{\star} A_l +A_l 
\stackrel{\scriptscriptstyle \bullet}{\star} \del_k \L\Big) \\
\stackrel{\scriptscriptstyle \bullet \scriptscriptstyle \bullet}{A_i}&=&
-\frac{1}{4}\t^{kl} \Big( \left\{\stackrel{\scriptscriptstyle \bullet}{A_k} 
\starcom \del_l A_i+F_{li} \right\} +\left\{A_k \starcom \del_l 
\stackrel{\scriptscriptstyle \bullet}{A_i}+
\stackrel{\scriptscriptstyle \bullet}{F_{li}} \right\} \\
&&+A_k \stackrel{\scriptscriptstyle \bullet}{\star} (\del_l A_i+F_{li})+
(\del_l A_i+F_{li}) \stackrel{\scriptscriptstyle \bullet}{\star} A_k \Big) .
\nonumber 
\ea
Notice that the equation for ${\del^n \L \over \del t^n}$ contains
${\del^{n-1} A_i \over \del t^{n-1}}$, while the equation for ${\del^n A_i 
\over \del t^n}$ depends only on ${\del^{k} A_i \over \del t^k}$, 
$k=0,\ldots,n-1$.
This means that the equations for $A$ are independent from those for
$\L$. We need to compute $A$ first and only afterwards we can substitute it in 
the expression for $\L$. If we use the homotopy operator, exactly the
opposite happens, we need the expression
for $\L^{(n)}$ first in order to obtain $A_i^{(n)}$.
If we insert the expressions (\ref{l1}) for 
$\stackrel{\scriptscriptstyle \bullet}{\Lambda}$
and (\ref{a1}) for $\stackrel{\scriptscriptstyle \bullet}{A_l}$ 
we obtain
\ba
\stackrel{\scriptscriptstyle \bullet \bullet}{\Lambda}&=&
\frac{1}{16}\t^{ij} \t^{kl}\Big(
\left\{ \left\{  \del_i \del_k \Lambda \starcom A_j \right\}+
 \left\{  \del_i \Lambda \starcom \del_k A_j\right\}A_l \right\} \nonumber\\
&&-\left\{\del_i \Lambda \starcom \left\{A_k \starcom \del_l A_j
+F_{lj} \right\} \right\} \label{lambda2} \\
&&+2i\left[\del_i \del_k \Lambda \starcom \del_j A_l\right]\Big).
\nonumber
\ea

\section{Ambiguities}

The solution (\ref{lambda2}) has to be compared to other known solutions
of the SW map at the second order, like e.g. \cite{GoHa} or \cite{JMSSW}. 
Before doing that, let us remark that the 
solutions of (\ref{swtl}) and (\ref{swta}) are not unique.
This has been commented on by a number of
authors \cite{BCPVZ,BCZ,GoHa,JMSSW,AsKi,BBG02}.
  
If we start with the structure equations 
(\ref{strl}), (\ref{stra})
\begin{eqnarray*}
s \L &=& i \L \star \L, \\
s A_i&=&\del_i \L -i [ A_i \starcom \L ],
\end{eqnarray*}
and consider a change in $\t$ by an amount $\delta \t$, then we see that
\be
\Delta ~\delta \L=-\ha \delta \t^{kl} \del_k \L \star \del_l \L~,
\ee
where the star product and the fields are at $t=1$ and where $\Delta$ is
the same as $\Delta_t$ for $t=1$.
Therefore, given a solution  $(\delta \L)_0$ of this equation,
\be
\delta \L =(\delta \L)_0+\Delta H
\ee 
is also a solution, because of the nilpotency of $\Delta$.

Similarly, for the gauge potential a change in $\t$ induces a change in $A_i$
determined by
\be 
\Delta ~\delta A_i=D_i \delta \L +\ha \delta \t^{kl} \left\{ \del_k A_i 
\starcom \del_l \L \right\}.
\ee
Therefore, given a solution $(\delta A_i)_0$ 
corresponding to $(\delta \L)_0$, the solution corresponding to 
$(\delta \L)_0+\Delta H$ is
\be
\delta A_i =(\delta A_i)_0+D_i H+S_i
\label{ambA}
\ee 
where $S_i$ is an arbitrary local functional of ghost number $0$
satisfying
\be
\Delta S_i=0.
\label{covamb}
\ee
This is a consequence of the fact that $\Delta$ commutes with the covariant
derivative: $D_i \Delta=\Delta D_i$.
The ambiguities determined by $H$ are of the form of a gauge transformation.

Due to the definition of $\Delta$ the condition (\ref{covamb}) means that
$S_i$ transforms covariantly
\be
s S_i=i [\L \starcom S_i].
\ee
This covariant ambiguity is of a different type from the gauge ambiguity.
It can be interpreted as a field dependent redefinition of the gauge potential.

The ambiguities of gauge type are an infinitesimal version of the 
Stora invariance \cite{BCZ} of the structure equations (\ref{strl}),
(\ref{stra})
\ba
\Lambda &\rightarrow& G^{-1} \star \Lambda \star G + i\,G^{-1} \star s G, 
\nonumber \\
A_i &\rightarrow&  G^{-1} \star A_i \star \,G + i\,G^{-1} \star \partial_i 
\,G ,
\label{st} 
\ea
where $G$ is an arbitrary local functional of ghost number $0$.

If we compare the solution to second order given in (\ref{lambda2}) for $t=0$
\be
\L^{(2)}=\ha \stackrel{\scriptscriptstyle \bullet \bullet}{\Lambda} \mid_{t=0}
\ee
with the solution $\L'^{(2)}$ found in \cite{GoHa} we see
that
\ba
\L^{(2)}-{\L'}^{(2)}&=&
\frac{1}{64} \t^{kl} \t^{mn} \Delta_0 \left(
\left\{\{D_m a_k+D_k a_m-f_{km}, a_n \}, a_l \right\} \right.\\
&&\left.-\left[[a_k,a_m],f_{nl}\right]\right), \nonumber
\ea
which is an ambiguity of the gauge type.

\section{Actions}

Until this point, we have discussed the deformation of gauge structures and
their representations in terms of Yang-Mills fields, without any reference to
the dynamics of the fields themselves.
To specify the dynamics, we must construct actions that are invariant under
the deformed gauge transformations

\be
\label{*gauge}
\begin{array}{ll}
s A_i=\partial_i \lam -i[A_i \starcom \lam],\\
s F_{ij}= -i[F_{ij} \starcom \lam].
\end{array}
\ee

The procedure is analogous to the construction of
commutative Yang-Mills theory. One arrives at the expression
\be
\label{noncommYM}
S^{YM}[A]=-\frac{1}{4}\int d^4 x \,Tr\, F_{ij}\star F^{ij},
\ee
where $F_{ij}$ is the noncommutative field strength given by (\ref{F}),
and the trace is the ordinary matrix trace in the appropriate representation.
The proof of the invariance of (\ref{noncommYM}) under (\ref{*gauge})
is based on the properties
\be                     
\label{quadraticaction}
\begin{array}{ll}
\int dx\, f \star g =\int dx\, fg =\int dx\, g \star f , \\
\int dx \,Tr\, M \star N = \int dx\, Tr\, N \star M,
\end{array}
\ee
the latter of which is valid for any pair of matrix valued functions,
when surface terms are ignored. Hence, the integral of the trace is
invariant under any cyclic permutation of its factors, also in the
presence of the star product.
Since the fields $A$ and $F$ are generally valued in the enveloping algebra,
we have to use the Seiberg-Witten map in order to make sense of
(\ref{noncommYM}) as a theory with a finite number of degrees of freedom,
namely those of $a_i$. To first order in the deformation parameter
$\theta$, we find
\begin{eqnarray}
S^{YM}=-\frac{1}{4} \int d^4 x \,Tr \,f_{ij} f^{ij} +\frac{1}{16} \theta ^{kl}
\int d^4 x \,Tr \,f_{kl}f_{ij}f^{ij} - {}\nonumber\\
\label{expandedYM}
-\frac{1}{2} \theta ^{kl}
\int d^4 x \,Tr \,f_{ik}f_{jl}f^{kl} + O( \theta ^2),
\end{eqnarray}
where $f_{ij}$ is the commutative field strength given by (\ref{f}).

We would like to remark that at the level of free fields there is
no difference between commutative and noncommutative theories,
because the properties (\ref{quadraticaction}) guarantee that the star product
disappears from any quadratic action. It is only when interaction terms are
present that the commutative and the noncommutative theories are in fact
different.
However, interaction terms are always present in the action (\ref{noncommYM}),
even if the gauge group is U(1), because of the star commutator term in the
expression (\ref{F}) for $F$.

In addition to the pure Yang-Mills theory, one can construct a noncommutative
version of any action with a gauge-invariance, simply by replacing each
ordinary product of functions with a star product, leaving the matrix
multiplication and the trace unchanged, and finally expanding each
noncommutative field by means of the Seiberg-Witten map associated with
the deformed gauge structure.
In particular, Yang-Mills theories with matter fields in various
representations have been considered by several authors.

Another gauge-invariant action that can be constructed in terms of gauge
potentials only is the Chern-Simons action in three dimensions.
Its deformed counterpart is obtained as described above and is
\be
\label{noncommCS}
S^{CS}_t[A]=\frac{1}{4\pi} \int d^3x \, \epsilon^{klm} Tr ( A_k \star
\partial_l A_m -\frac {2}{3} i A_k \star A_l \star A_m ),
\ee
where the subscript $t$ refers to the parameter of the evolution equation
described in section 3.

If one were to expand (\ref{noncommCS}) by means of the Seiberg-Witten map,
one would find that it is in fact identical to the undeformed
action \cite{GrSi}.
In other words
\be
\label{identical}
S^{CS}_1[A]=S^{CS}_0[a].
\ee
This can be proven to hold at all orders in the deformation parameter,
by showing instead
\be
\label{derivative}
\frac{d}{dt}S^{CS}_t[A]=0,\; \forall t.
\ee
The total $t$-derivative is computed using
\be
\label{der*}
f \stackrel{\scriptscriptstyle \bullet}{\star} g = \frac{i}{2} \theta ^{kl} 
\partial_k f \star \partial_l g
\ee
and the evolution equation for $A$
\be
\label{evolution}
\stackrel{\scriptscriptstyle \bullet}{A_k}=-\frac{1}{4} \theta^{rs} 
\{A_r \starcom \partial_s A_k + F_{sk} \}.
\ee             
In this context, it is worth noting that the WZW model in two dimensions
shares the same property, namely the identity of the actions for the
commutative and the noncommutative version \cite{MoScha}, and that the 
WZW model in two dimensions is related to the Chern-Simons action in three.

\section{Concluding remarks}

In this paper we have limited ourselves to gauge theories of the Yang-Mills
type and have based our analysis on the structure equations (\ref{unstrl}) and 
(\ref{unstra}) (which involve BRST transformations) and their deformation.
This formulation is sufficient for Yang-Mills theories, but for gauge theories
with reducible gauge transformations, such as theories with gauge potentials
which are differential forms of degree higher than one, it is appropriate
to use the antifield formalism of Batalin and Vilkoviskii (BV). The deformation
of the gauge structure should then be studied by defining generalized
Seiberg-Witten maps in the context of the BV
formalism \cite{BGH,BBG01,BBG02}. The use of the master equation couples
intimately the gauge transformations and the dynamics, i.e. the
action functional.

The existence of the SW map, together with the understanding of its
ambiguities, can be interpreted as a kind of ``rigidity'' of the structure
of the gauge group, analogous to the rigidity of semisimple Lie algebras under
smooth deformations of the structure constants: the structure constants
can be brought back to their original values by performing a linear
transformation on the generators. In the case of gauge groups the deformed
structure equations can be reduced to the undeformed equations by expressing
the deformed fields (e.g. $A_i$ and $\Lambda$) as local functionals of the
undeformed fields ($a_i$ and $\l$). Strictly speaking, we have discussed
only infinitesimal gauge transformations in a context in which only the
space-time coordinates are deformed. Thus, we have ignored all questions
for which the topology of the gauge group may be relevant when the gauge
fields are quantized \cite{Pol}.

As explained in the introduction, throughout this paper we have considered
the case of $\theta^{ij}$ independent of $x$. Techniques of deformation
quantization are available for an $x$-dependent Poisson tensor (see,
e.g. \cite{CaFe1,CaFe2}, and references therein, where general coordinate
transformations for quantized
coordinates are also studied). It would be interesting to extend to that
case the results described in the previous sections.

Recently, several authors have pointed out the analogy of the  Weyl-Moyal
star product with the associative, noncommutative star product which enters
in the formulation of Witten's bosonic open string field
theory \cite{Wi,Ba,BaMa04,BaMa08,DLMZ,Bel}. It would be
interesting if methods of deformation quantization developed in the context
of the Seiberg-Witten map, would turn out to be useful in string field theory.

\section*{Acknowledgements}

We are very grateful to S.~Schraml and A.~Weinstein for many fruitful
discussions. This paper is a written version of a talk given by B.~Zumino
at the Workshop on ``Continuous Advances in QCD'', Minneapolis, Minnesota,
May 17-23, 2002.
This work was supported in part by the Director, Office of
Science, Office of High Energy and Nuclear Physics, Division of High 
Energy Physics of the U.S. Department of Energy under Contract 
DE-AC03-76SF00098 and in part by the
National Science Foundation under grant PHY-0098840. B.L.C. is supported
by the DFG (Deutsche Forschungsgemeinschaft) under grant CE~50/1-2.

\end{document}